# Storage Solutions for Big Data Systems: A Qualitative Study and Comparison

Samiya Khan[a,1], Xiufeng Liu[b], Syed Arshad Ali[a], Mansaf Alam[a,2]

[a]*Jamia Millia Islamia, New Delhi, India*
[b]*Technical University of Denmark, Denmark*

---

**Highlights**

- Provides a classification of NoSQL solutions on the basis of their supported data model, which may be data-oriented, graph, key-value or wide-column.
- Performs feature analysis of 80 NoSQL solutions in view of technology selection criteria for big data systems along with a cumulative evaluation of appropriate and inappropriate use cases for each of the data model.
- Classifies big data file formats into five categories namely text-based, row-based, column-based, in-memory and data storage services.
- Compares available data file formats, analyzing benefits and shortcomings, and use cases for each data file format.
- Evaluates the challenges associated with shift of next-generation big data storage towards decentralized storage and blockchain technologies.

---

**Abstract**

Big data systems' development is full of challenges in view of the variety of application areas and domains that this technology promises to serve. Typically, fundamental design decisions involved in big data systems' design include choosing appropriate storage and computing infrastructures. In this age of heterogeneous systems that integrate different technologies for optimized solution to a specific real-world problem, big data system are not an exception to any such rule. As far as the storage aspect of any big data system is concerned, the primary facet in this regard is a storage infrastructure and NoSQL seems to be the right technology that fulfills its requirements. However, every big data application has variable data characteristics and thus, the corresponding data fits into a different data model. This paper presents feature and use-case analysis and comparison of the four main data models namely document-oriented, key-value, graph and wide-column. Moreover, a feature analysis of 80 NoSQL solutions has been provided, elaborating on the criteria and points that a developer must consider while making a possible choice. Typically, big data storage needs to communicate with the execution engine and other processing and visualization technologies to create a comprehensive solution. This brings forth second facet of big data storage, big data file formats, into picture. The second half of the research paper compares the advantages, shortcomings and possible use cases of available big data file formats for Hadoop, which is the foundation for most big data computing technologies. Decentralized storage and blockchain are seen as the next generation of big data storage and its challenges and future prospects have also been discussed.

---

Corresponding authors.
E-mail addresses: [1]samiyashaukat@yahoo.com (Samiya Khan), [2]malam2@jmi.ac.in (Mansaf Alam)

**1.0 Introduction**

Data is processed to generate information, which can later be used for varied purposes. Data mining and knowledge discovery are two fields that have been actively working towards deriving useful information from raw data to create applications that can make predictions, identify patterns and facilitate decision making [1]. However, with the rise of social media and smart devices, data is no longer a simple dataset that traditional tools and technologies can handle [2].

Digitization and rising popularity of modern technologies like smart phones and gadgets has contributed immensely towards 'data deluge'. Moreover, this data is not just high on volume, but it also includes data of varied kinds that is generated on a periodic basis. The biggest challenge in dealing with this 'big data problem' is that the present or traditional systems are unable to store and process data of this kind. Therefore, this gave rise to the need for scalable systems that can store varied forms of data and process the same to generate useful analytical solutions [3].

The present era can rightly be called the era of analytics where organizations are tapping the business potential of data by processing and analyzing it. A plethora of technologies are available for this purpose and organizations are smoothly drifting towards heterogeneous environments, which include data stores like HBase [4], HDFS [5] and MongoDB [6], execution engines like Impala [7] and Spark [8], and programming languages like R [9] and Python [10].

Big data storage [11] is a general term used for describing storage infrastructures designed for storage, management and retrieval of data that is typically large in volume, high in velocity and diverse in variety. In such infrastructures, data is stored in such a manner that its usage, processing and access become easier. Moreover, such infrastructures can scale as per the requirement of the application or service.

The primary task of big data storage is to support input and output operations on stored data in addition to storage of a large number of files and objects. Typically, the architectures used for storage of big data include a cluster of network-attached storage, pools of direct attached storage or storage based on object storage format [11]. Computing server nodes are used at the heart of these infrastructures in order to provide support for retrieval and processing of big data. Most of these storage infrastructures provide support for big data storage solutions like Hadoop [12] and NoSQL [13].

The storage needs of a big data problem are influenced by many factors. Fig. 1 recapitulates the requirements from a big data storage system, which include –

- Scalability
  Scalability is undoubtedly one of the fundamental requirements in view of the ever-growing size of data. Any solution made for big data must be able to accommodate the growing data in an optimal manner.
- Availability
  Considering the fact that most big data solutions require real-time analysis of data and its visualization, the allowable time in which data must be accessed is extremely low. Moreover, data needs to be accessed in a frequent and efficient manner, making availability a crucial system requirement.
- Security
  Big data solutions may make use of organization-specific data. One of the biggest concerns of organizations in the adoption of such solutions is the security of their data.
- Integration
  There may be a need for big data solutions to interact with other technologies and applications. Therefore, big data solutions must be able to integrate with these solutions to create a complete application.

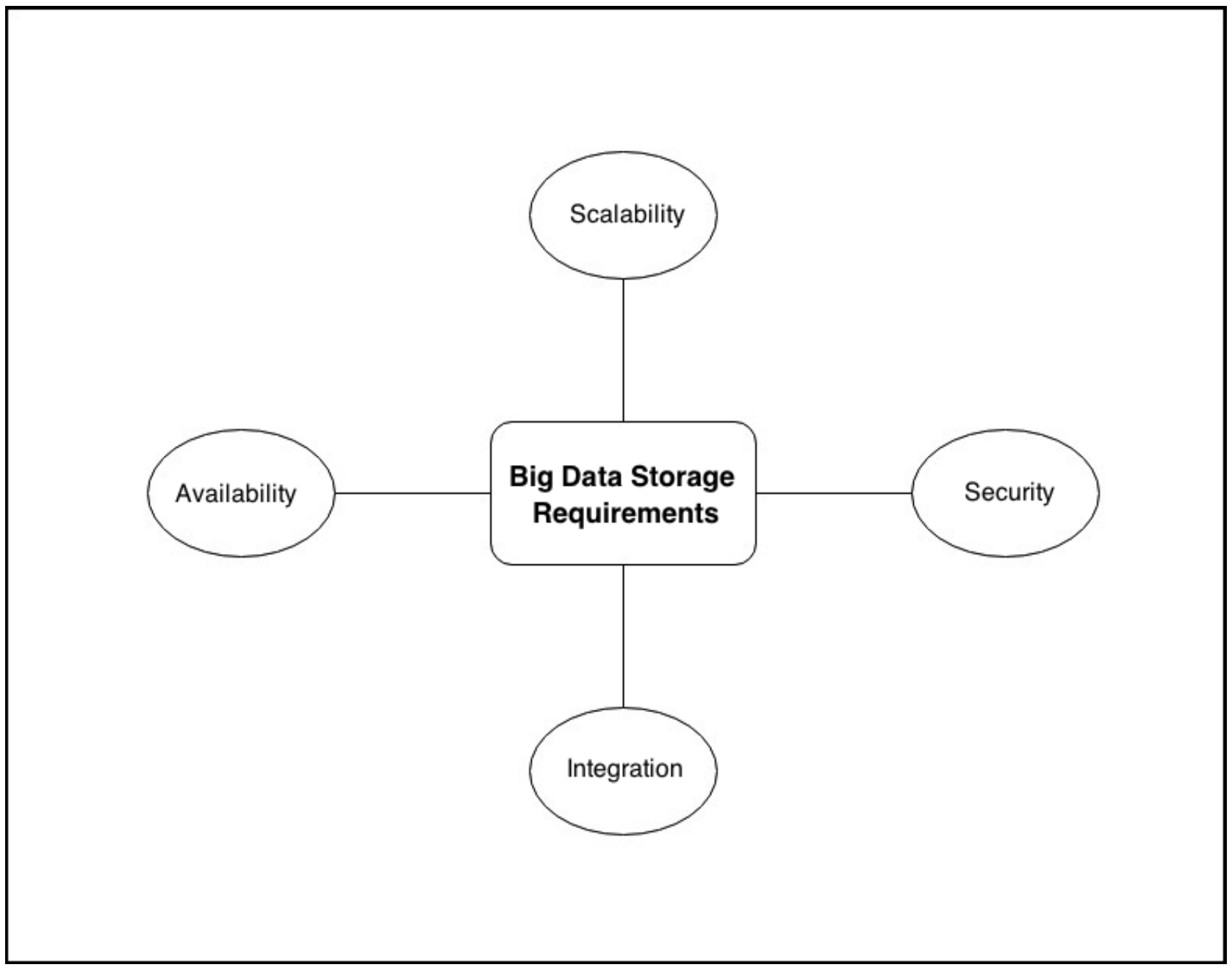

Fig. 1. Storage Requirements of a Big Data System

Technologies or solutions must be chosen on the basis of the specific requirements of a business or application. The available big data technologies offer different degrees of performance, security, and data capacity and integration capabilities [14]. Therefore, if the requirements are clear and precise, choosing a solution or combination of solutions to suffice the needs should not be difficult. This research paper performs a comparative study of 80 NoSQL solutions available for use in big data systems and elaborates on how requirements must be analyzed to determine the best solution for a concerned application.

Understandably, sustainability of heterogeneous environments requires data sharing between multiple processes, which can be performed using two possible ways. The first method makes use of APIs provided by independent solutions to retrieve data from them. For instance, if Drill [15] and HBase [4] are used together, Drill shall use the HBase API for data reads. Although, data sharing can be performed in real-time using this approach, it suffers from some potential issues [16].

The overheads for deserialization and serialization are very high and may cause serious bottleneck issues for applications like workflows. Moreover, there is no standardization in how the data should be represented. Developers perform custom integrations using their system and other systems. In order to manage these issues, another method for data sharing may be used, which enables sharing by means of common file formats that are optimized for high performance in analytical systems [17]. Owing to this, big data file formats form an important facet of big data storage optimization and are discussed in the later parts of this paper. This research work classifies data file formats and compares the advantages and disadvantages of each to provide a comprehensive comparison for use-case matching.

Recent surveys on NoSQL technologies [18, 19, 20] and data file formats [21, 22] discuss these two storage facets in isolation, which cannot be done in consideration of the fact that big data storage and its optimization need to be dealt with, in entirety, in this age of heterogeneous systems development that integrate different solutions to get the desired functional and performance results. Moreover, the coverage of this study is largest among available research papers. It covers 80 NoSQL solutions and reviews big data file formats comprehensively, which is not present in available literature. The rest of the research paper has been organized in the manner discussed below. Section 2 describes NoSQL and its characteristics. It aims to answer as to how NoSQL solves the multiple issues presented by the big data problem to traditional systems.

Big data storage models have been discussed in Section 3. The research paper provides a classification of popular NoSQL solutions on the basis of data model-based categorization. The different categories are compared to provide an analysis of matching use-cases, to facilitate decision making in this domain. Section 4 discusses big data file formats in detail. All the available formats are divided into five categories and are compared to provide the pros and cons of using each of them. Moreover, best-suited applications for each of these formats are also discussed. Most of the available solutions are cloud-based, but big data storage is still not decentralized enough to provide optimum results. Future trends in big data technology deals with decentralized storage and blockchain. These prospects have been discussed in Section 5. Lastly, the research paper concludes and summarizes the findings in Section 6, providing insights for future work in this field.

## 2.0 NoSQL – A Solution for Big Data Storage Issues

The idea behind the development of relational databases was to provide a data storage approach that makes use of structured query language or SQL [23]. The introduction of these databases dates back to the 1970s when data schemas were not as complicated as they are today. Moreover, storage was expensive and not all data was considered for archival. With the rise in social media platforms, the amount of data being stored about events, objects and people has risen exponentially. The use of data in this time and age is not just limited to data archival, but it also extends to frequent data retrieval and processing, in order to serve purposes like generation of real-time feeds [24] and customized advertisements [25], in addition to many others.

Owing to the complexity of information being processed and the need to treat multiple, of the order of hundreds, database requests just to answer a single API request or render a webpage, the demands from modern database systems are ever-increasing. Some of the key drivers in this domain are the need for interactivity, increasing complexity and ever-evolving networks of users [26]. In order to serve these growing demands, sophisticated deployment strategies and improved computing infrastructure [188] are being put to use. With that said, single server deployments are expensive and highly complex, which has caused a drift towards the use of cloud hardware [27] for this purpose. Besides this, the use of agile methods has also reduced the development and deployment time [28], allowing quicker response to user needs.

It would not be wrong to state that relational databases were not created to manage the agility and scalability requirements of modern-day systems. Moreover, they are also not equipped to work with the cloud and take optimum advantage of its cheaper storage and processing capabilities. These shortcomings can be addressed using two main technical approaches, which have been discussed below –

- Manual Sharding

  In order to make use of the distributed paradigm, tables need to be segmented into smaller units, which must then be stored across different machines. This process of splitting is called manual sharding [29]. However, this functionality is not available in a traditional database and needs to be implemented by the developer. Moreover, the storage of data on each instance is performed in an anonymous mode.

  It is the responsibility of the application code to segment data, store it in a distributed manner, and perform query management and aggregate results to be presented to the user. Additional code shall be required for supporting data rebalancing, performing join operations, handling of resource failures and replication. It is crucial to mention that manual sharding may downgrade some of the benefits of relational databases like transactional integrity.

- Distributed Cache

  Caching [30] is a commonly used process, which is primarily employed for improving the read performance of a system. It is noteworthy that the use of a cache has no impact on the write performance and is capable of adding substantially to the complexity of the overall system. Therefore, if the requirements from the system are read-intensive, then the use of distributed cache must be considered. On the other hand, write-intensive or read/write intensive applications do not require a distributed cache [31].

NoSQL databases [13] are known to mitigate the challenges associated with traditional databases. In addition, they also unleash the true power of cloud by making use of commodity hardware, which reduces the cost, and simplifies deployment, making the life of a developer much easier as there is no need to maintain multiple layers of cache anymore.

Some of the advantages of NoSQL solutions over traditional databases are as follows –

- Scalability

  NoSQL allows systems to scale out horizontally [13]. Moreover, this can be done quickly without affecting the overall performance of the system with the help of cloud technologies. Scaling traditional databases require manual sharding that involves high costs and complexity. On the other hand, NoSQL solutions offer automatic sharding, reducing complexity as well as cost of the system [13].

- Performance

  As mentioned previously, NoSQL systems can be scaled out as required. With the increase in the number of systems, the performance of a system is also correspondingly improved. The fact that these systems involve automatic sharding means that the overhead associated with the same is also eliminated, which further contributes to the improved performance of the system.

- High and Global Availability

  Relational databases depend on primary and secondary nodes to fulfill the availability requirements. This not only adds to the complexity of the system, but it also makes the system moderately available. On the contrary, NoSQL solutions make use of master-less architecture and data is distributed across multiple systems [13]. Therefore, even upon the failure of a node, the availability of the application remains unaffected for read as well as write operations.

  NoSQL solutions offer data replication across resources [13]. Consequently, user experience is consistent irrespective of the location of the user. Moreover, it also plays a significant role in reducing latency with the added advantage of shifting the developer's focus from database administration to business primacies.

- Flexible Data Modeling

  It is possible to implement fluid and flexible data models in NoSQL [13]. This allows developers to implement query options and data types that befit the application instead of those that suit the schema. In the process, the interaction between database and application is simplified, making this approach a better option for agile development.

NoSQL is an umbrella term used to describe a plethora of technologies, all of which entail the following common characteristics [32] –

**2.1 Dynamic Schemas**

Relational databases [33] have an inherent requirement to create schemas in advance. Data is added to the database only after this requirement is fulfilled. For instance, if a system needs to store employee data like name, department, age, gender and salary, then the table created for the same must have the corresponding schema.

Such a requirement is unfit for agile development environments, as the fields of data might need to be changed over time. A new requirement may be added, as part of iteration, and subsequently, the schema may have to be altered. This is a time-consuming task if the database is large. As a result, the database may have to be shut for any use for a considerable amount of time to make required changes. Moreover, if the development process requires several iterations, the database may have to be shut rather frequently for significant amounts of time. Evidently, relational databases are inappropriate for storing data that are large, unstructured and unknown [33].

NoSQL satisfies this requirement since it has no predefined schemas. Moreover, data insertion does not require the developer to define a schema well in advance. As a result, changes to the data structure and data can be made in real-time without the need to shut the database for any other use [32]. There are several advantages of using this approach. Apart from the fact that it reduces administrator time, such an approach also reduces the time required for development and simplifies the process of code integration.

**2.2 Auto-Sharding**

Relational databases are structured in such a manner that they need to have a server that controls the rest of the systems to provide reliability and availability requirements of a database solution. Therefore, such a system can only support vertical scaling [32], which is not just expensive, but it leads to creation of small number of points of failure. Besides this, it also places a limit on the amount of scaling that a system can support.

In view of the system requirements, a database solution must support horizontal scaling [13]. Therefore, it must be possible to add servers to the ensemble and get rid of the limitation that focuses on testing the capacity of a single server. Cloud Computing offers the best solution in this regard by providing on-demand services and unlimited scaling capacity [34]. So, the system no longer needs to rely on one server to fulfill its needs. Another important facet of using the Cloud is its inbuilt database administration. Moreover, the developer no longer needs to create complex platforms and can simply focus on writing the application code. Lastly, the use of Cloud-based, multiple servers cost significantly lesser than a high-capacity, single server.

In order to perform sharding of a database spanning across multiple servers, complex arrangements to make multiple servers act a single system, need to be put in place. On the other hand, NoSQL databases support auto-sharding. In other words, the database automatically distributes data across multiple systems without the need for the administrator to be aware of the server pool composition. Load balancing [35] for data and query are also automatically performed by the system. This allows the system to offer high availability. As and when the server goes down, it can be conveniently replaced and operations remain unaffected.

**2.3 Automatic Replication**

Replication is performed automatically for any NoSQL system [32]. Therefore, the system can recover from disasters rather easily, also allowing high degrees of availability. From the developer's point of view, he or she no longer needs to cater for these facets of development in the application code.

**2.4 Integrated Caching**

The integrated caching abilities [32] of NoSQL systems are rather well equipped and most of the frequently used data is kept in the system memory to ensure quick access. Therefore, there is no need to maintain multiple caching layers at the application level.

## 3.0 Big Data Storage Models

NoSQL is a technology that is developed to counter the issues presented by relational databases, which is implemented in multiple ways by different models. Common characteristics of NoSQL models include efficient storage, reduced operational costs, high availability, high concurrency, minimal management, high scalability and low latency [36]. NoSQL solutions have been classified using multiple criteria.

Yen [37] provided a detailed classification of NoSQL solutions dividing them into nine categories, which include Wide Columnar Store, Document Store, Object Database, Tuple Store, Data Structures Server, Key Value Store, Key-Value Cache, Ordered Key Value Store and Eventually Consistent Key Value Store. Another taxonomical study was performed by North [37], which gave a comprehensive classification and included cloud-based solutions as well for the analysis. Solutions have been classified under six categories namely Entity-Attribute-Value Data Stores, Amazon Platform Column Stores, Key-Value Data Stores and Distributed Hash Table Document Stores.

Catel [39] and Leavitt [40] proposed a data model-based classification. Catel [39] divides NoSQL solutions into three categories namely Key Value Stores, Document Stores and Extensible Record Stores. On the other hand, Leavitt [40] proposes the use of three categories namely document-based, key value stores, column-oriented stores. Scofield [41] gave the most accepted categorization scheme by classifying databases into relational, graph, document, column and key value stores. This research paper uses the above-mentioned basis for classification and covers document-oriented stores, graph data model, key value store and wide column store in the following sections.

### *3.1 Document-Oriented Data Model*

A NoSQL database that uses documents for storage and retrieval of data is called document-oriented database [42]. Other names of this NoSQL data model are document database [43] and document store [44]. It is primarily used for management of semi-structured data because of its flexibility and support for variable schema. As mentioned previously, document databases use documents for its working. These documents may be in PDF or word format. However, blocks of JSON and XML are the more commonly used document formats. A relational database contains columns, which are described by their names and data types. On the contrary, in case of document databases, data type description and value for the concerned description are provided in a document [42]. The structure of different documents making up a database may be similar or different in structure. Evidently, there is no need to alter the schema for adding data to the database.

Documents are grouped together to form a structure called collection [42]. There may be multiple collections in a database. This structure is similar in functioning to the table, which is present in a relational database [33]. Document-oriented databases provide a mechanism to execute queries on collections and retrieve documents that satisfy the attribute requirements. There are several advantages of using this approach, which include –

- Most of the growing data comes from IoT devices, social media and Internet. However, this data does not fit into standard application data models. Document-oriented databases offer flexible data modeling [42], in contrast to relational databases that force applications to fit data into existing models irrespective of their needs.
- The write performance of document-oriented databases is better than conventional systems [45]. In order to make a system available for writing, the database can compromise on data consistency as well. Therefore, even if a system fails and replication takes longer than expected, the write operation will be fast.
- The indexing features and query engines of databases available in this category are known to be fast and efficient [42]. Therefore, they offer faster query performance.

### *3.2 Graph Data Model*

This NoSQL data model is tailor made to support storage and processing of voluminous data, which may be semi-structured, structured or unstructured, in type. Therefore, data can be accessed and acquired from different sources. As a result, Graph data model [46] is popularly used in social media [47] and big data analytics [48]. It is noteworthy that relational databases were developed for storing structured information available in and generated by enterprises. Therefore, schema of the data to be stored is available beforehand. On the contrary, data generated by IoT (Internet of Things) and social media is unstructured. Moreover, it is being generated in real time.

Graph databases [49] are a good option for storing unstructured data generated by such diverse sources at high velocity. There is no need to define a schema before storing data, which makes the database rather flexible. Besides this, graph databases are cost-effective and dynamic when it comes to integration of data coming from different sources [49]. Moreover, graph databases are better equipped to handle, store and process high-velocity data as compared to relational databases.

Aforementioned applications like social media analytics and IoT-based analytical solutions [187] require the base technology to integrate data coming from heterogeneous sources and establish links between the different datasets created. Application data of this kind can best be handled using a semantic graph database or RDF triplestore [50], which focuses on relationships between different elements of the database and generate analytics on this basis. These graph databases are primarily used for real time analytics because of their ability to handle large datasets, without the need to define a schema in advance.

The benefits of using semantic graph database can be summarized as follows –

1. Integration of inbound data from different sources is limited when the schema needs to be defined before adding data because the addition of a new source might require a change in schema, which is both time-consuming as well as complicated. In databases where there is no such need, data integration is limitless, simple and cost-effective [51].
2. Semantic graph databases offer an additional support to ontologies or semantically rich data schemas [51]. Therefore, organizations can create logical models in any way they desire.
3. Semantic graph databases use international standards for data representation on the web [51]. This results in easier integration and sharing of data. Uniform Resource Identifier (URI) [52] is one of the standards used for data representation in semantic graph databases. URI is a unique ID, which is used to distinguish between linked entities. The presence of such a clear approach for entity identification makes access and search easier, making the approach cost-effective. Moreover, it also makes data sharing easier as far as mapping data to Linked (Open) Data is concerned. In addition, challenges like vendor lock-in can be avoided.

### *3.3 Key-Value Data Model*

The most flexible type of NoSQL database is Key-Value Store [53], which implements schema-less policy by having no schema and making the data value opaque. The data value can store strings, numbers, images, binaries, counters, XML, JSON, HTML and videos, in addition to many others [53]. The stored values can be accessed with the help of a key. The flexibility of the database is manifested in the fact that the application controls the data value, completely. Key benefits of key-value stores are as follows –

1. The database does not force the application to structure its data in a specific form. Therefore, the application is free to model its data in accordance with the requirements of the use case.
2. Objects can simply be accessed with the help of the key assigned to the object. When using this database, there is no need to perform operations like union, join and lock on objects [53], which make this data model, most efficient and high performing.
3. Most of the available key-value databases allow scale out as and when the demand for the same arises. Moreover, this can be done using commodity hardware without the need for any redesigning.
4. Providing high availability is much easier and uncomplicated with key-value stores. The distributed architecture and master-less configuration of some of the available databases of this type ensures higher resilience [53].
5. The design of these databases is such that it is simple to add and remove capacity. Moreover, these databases are better equipped to deal with network failures and hardware malfunctions [53], lowering the downtime considerably.

### *3.4 Wide-Column Data Model*

Wide Column Stores [54] have columns and column families, as base entities. Facts or data are grouped together to form columns, which are further organized in the form of column families that are constructs similar to tables in relational databases. For example, data about an individual like name, account name and address are facts about the individual and can be grouped together to form a row in a relational database. On the contrary, same facts are organized in the form of columns in a wide-column store and each of the columns includes multiple groups. Therefore, a single wide-column can store data equivalent to the same stored by many rows in a relational database. Other names of such databases include column-oriented DBMS [55], columnar databases [56] and column families [57].

Key advantages [54] of using wide column store databases include –

1. Partitioning and data compression can be performed efficiently using wide column store databases.
2. Aggression queries like AVG, SUM and COUNT can be performed effectively and efficiently because of the inherent structure of this database.
3. This database type is highly scalable and well suited for massively parallel processing (MPP) systems.
4. Tables with huge amounts of data can be loaded and queried in almost no time, making the response time of the database relatively low.

### *3.5 Choosing a NoSQL Solution for a Big Data System*

One of the major technological decisions to be made while designing a big data application include selecting a NoSQL solution. For this, two things need to be specifically kept in mind. The right data model for an application depends on the type of data that needs to be dealt with. A classification of NoSQL databases on the basis of their data model is provided in the previous section. The success of a big data application can be greatly impacted by a mismatch in the data model of the application and that of the chosen NoSQL solution. Another important consideration in choosing a NoSQL solution is the scalability requirement. It is critical to understand that there are some NoSQL solutions that can scale well like Cassandra whereas others may be memory-based and fail to scale across machines.

As mentioned previously, the right data model for an application depends on the data that it is expected to deal with. For instance, if the application’s data can be represented in the form of a graph, then the graph model is most appropriate data model for the application. Each data model best suits to a specific set of applications and requirements. This section discusses the selection criteria for deciding which big data model befits a particular case study in view of its domain model, data access patterns and use cases.

Document databases work around documents. Therefore, a document is the basic atomic unit of storage in such databases. Any domain model that allows splitting and partitioning of its data across documents can use a document database. Some common case studies include CMS, blog-software and wiki-software [58]. However, when considering this data model, you may come across use-cases where a relational model may be just as good an option to use as the non-relational database.

Key-value stores are commonly used data models, preferred for application areas surrounding data like user profile, emails, blog/article comments, session information, shopping cart data, product reviews, product details and Internet Protocol (IP) forwarding tables [59], in addition to many others. It is crucial to understand that a key-value store can be used to store complete webpages [60]. In this case, URL can be used as the key, and webpage content, as value. However, other data models may be better suited for this purpose if the application requires so.

Graph databases offer an effective way to manage and combine data. Enterprise data is typically linked and graph databases for their storage ensure easier management of content. Moreover, personalization can also be achieved in a simpler manner. In addition to this, the concept of connected world, which has particularly picked up pace after the rise of social media and IoT, can take advantage of the fact that graph databases allow integration of heterogeneous, interlinked data from different sources.

Wide-column stores form the last category of big data models. These databases are deemed most appropriate for distributed systems [61]. In other words, if the data available is large and can be split across machines, then a wide-column store database can be extremely useful. Some of the primary advantages of using this database is reduced query time for some queries. However, this point must be clearly investigated before a decision in favor of such a solution is taken. For some queries, the time may be same or higher than that offered by conventional RDBMS solutions [62]. The use-cases for the four big data models have been summarized in Table 1.

Table 1: Use Cases for Different Big Data Models

| Big Data Model | Preferred for Use Cases | Not Preferred for Use Cases |
|---|---|---|
| Document-Oriented | 1. Content management systems<br>2. E-commerce platforms<br>3. Blogging platforms<br>4. Analytics platforms | 1. Applications requiring complex search queries.<br>2. Applications requiring complex transactions with multiple operations. |
| Key-Value | 1. Storage of user preferences.<br>2. Maintenance of user profiles that don’t have a specific schema.<br>3. Storage of session data for users.<br>4. Storage of shopping carts’ data for multiple users. | In scenarios where –<br>1. Specific data value needs to be queried.<br>2. Multiple unique keys need to be worked upon.<br>3. Frequent update of a part of the value.<br>4. Data values have established relationships with each other and the application requires exploitation of the same. |
| Graph | 1. Network and IT operations<br>2. Graph based searches<br>3. Social networks<br>4. Fraud detection | Such a model is inappropriate for any application for which the data cannot be modeled as a graph. |
| Wide-Column | 1. Blogging platforms<br>2. Content management systems<br>3. Counter-based systems<br>4. Applications with write-intensive processing | 1. Application requires complex querying.<br>2. Application has varying patterns of queries.<br>3. In scenarios where the database requirement is not established, the use of such a store must be avoided. |

While discussing the applicability of NoSQL solutions to real-world problem, it is important to mention CAP Theorem [63]. This theorem introduces the concept of Partition Tolerance (P), Availability (A) and Consistency (C) for distributed systems and states that all these three characteristics cannot be ensured by a solution simultaneously. Consistency is a characteristic that ensures that all the nodes of the distributed system must read the same value of data at all times. If a change in data value is made, then the change must be consistent for all nodes. However, if the change results in an error, then a rollback must be performed to ensure consistency.

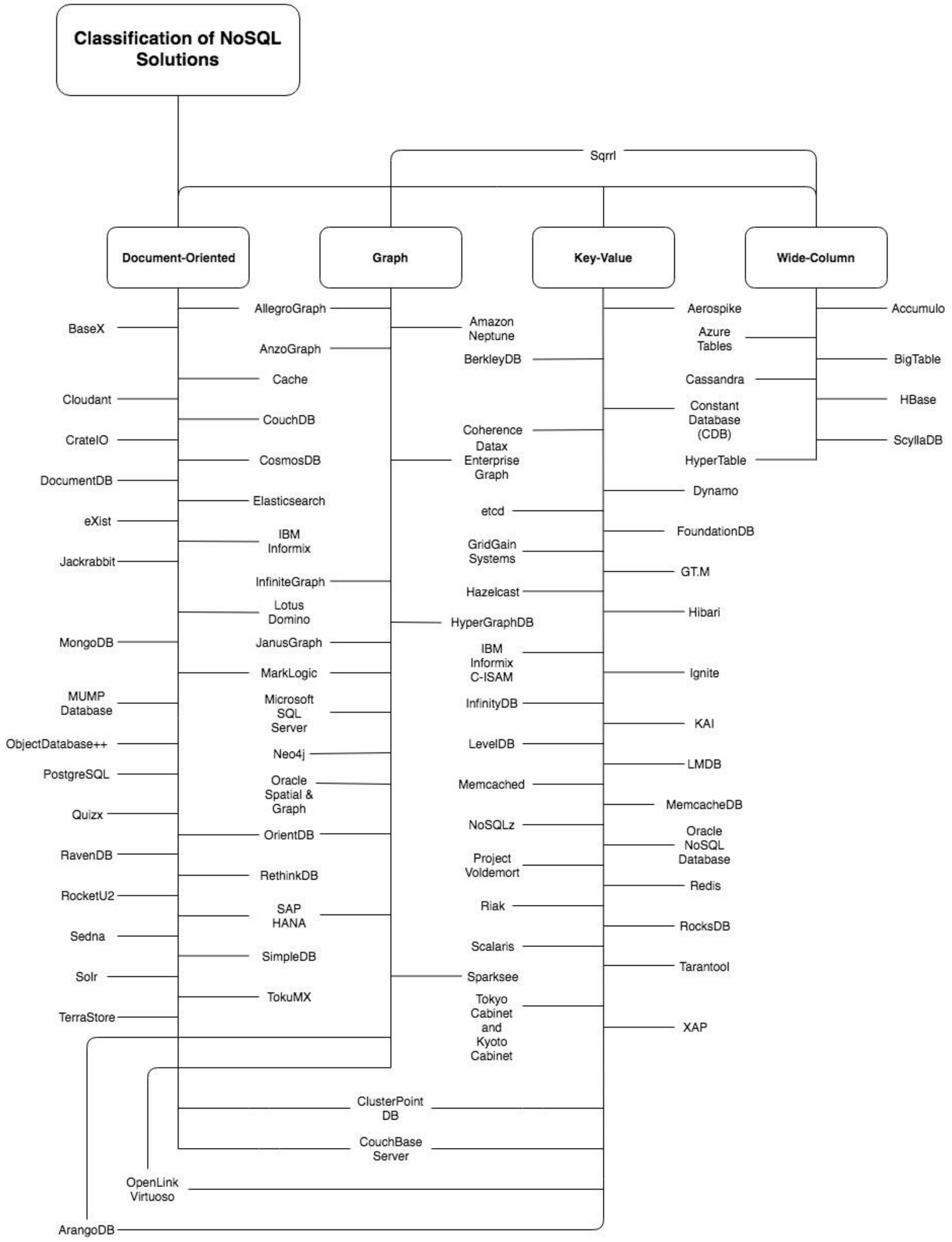

Fig. 2. Data Model-Based Classification of NoSQL Solutions

Availability defines the operational requirement of the system that ensures that as and when a request is made to the system by a user, it must be respond to it despite its state. Partition tolerance refers to a system's ability to operate despite failure of a partition and message loss. It can also be described as the ability of a system to operate irrespective of network failure. Different database solutions and their CAP status has been described in the following sections. It has been stated that no distributed system can possess all the three characteristics. On the basis of this assertion, NoSQL systems can be CA (Consistent-Available), AP (Available-Partition Tolerant) or CP(Consistent-Partition Tolerant) [64].

One of the biggest challenges in the use of NoSQL is the use of an appropriate data model. The use of an inappropriate model can affect system performance significantly, which makes this a crucial technological decision. Another important technological decision in this case is choosing the right distribution model. Scaling of read operations is supported by master-slave architecture. However, if scaling of both read and write operations is desired, then peer-to-peer architecture is a better option. The use of NoSQL databases also has some security issues that must be considered and mitigated before a solution can be developed and deployed using the same. Fig. 2 illustrates data model-based classification of NoSQL solutions. A detailed description of implementations along with the supported data models is provided in Table 2. The letter Y indicates support for the corresponding data model while N suggests no support for the concerned data model.

Table 2: Comparison of Solutions Based on Different Data Models

| **S. No.** | **Implementation** | **Supported Data Model** | | | | **Key Features** |
|---|---|---|---|---|---|---|
| | | Document-Oriented | Graph | Key-Value | Wide-Column | |
| 1. | AllegroGraph [65] | Y | Y | N | N | 1. Proprietary<br>2. Supports RDF, JSON and JSON-LD<br>3. Provides Multi-Master Replication<br>4. Supports ACID Transactions, two-phase commit and full-text search |
| 2. | Accumulo [66] | N | N | N | Y | 1. Free<br>2. Scalable and distributed<br>3. It is built over and above Hadoop [8], Thrift [74] and Zookeeper [107]<br>4. Provides cell-level security and mechanisms for server-side programming |
| 3. | Aerospike [67] | N | N | Y | N | 1. Free<br>2. Highly scalable<br>3. Flash-optimized, in-memory<br>4. Reliable and consistent<br>5. Used for applications like dynamic web portals, user profiling and fraud detection |
| 4. | Amazon Neptune [68] | N | Y | N | N | 1. Proprietary<br>2. Fully managed database<br>3. Provided as a web service<br>4. Supports RDF and property graph models<br>5. Supports SPARQL [196] and TinkerPop Gremlin [197] query languages |
| 5. | AnzoGraph [69] | N | Y | N | N | 1. Proprietary<br>2. Massively parallel<br>3. Graph Online Analytics Processing (GOLAP) database<br>4. Supports SPARQL [196] and Cypher [145]<br>5. Originally designed to analyze semantic triple data interactively |
| 6. | ArangoDB [70] | Y | Y | Y | N | 1. Free<br>2. Supports multiple database models with a single core<br>3. Possesses unified query language called ArangoDB Query Language (AQL) |
| 7. | Azure Tables [71] | N | N | N | Y | 1. Proprietary<br>2. Provisioned as a service where storage of data is allowed into collections that can be partitioned. Data is accessed by means of primary and partition keys. |
| 8. | BaseX [72] | Y | N | N | N | 1. Free<br>2. Provides support for JSON, XML and binary formats<br>3. Implements master-slave architecture<br>4. Provides support for concurrent structural and full-text update/search |
| 9. | BerkeleyDB [73] | N | N | Y | N | 1. Free (with commercial versions also available)<br>2. High performing and scalable<br>3. Supports complex data management<br>4. Most appropriate for applications requiring embeddable database. |
| 10. | BigTable [75] | N | N | N | Y | 1. Proprietary<br>2. High performance<br>3. Provides data compression |
| 11. | Cache [76] | Y | N | N | N | 1. Proprietary<br>2. Data is stored in multi-dimensional arrays. Therefore, structured data that is hierarchical in nature can be stored.<br>3. Commonly used for business and health-related applications |
| 12. | Cassandra [77] | N | N | N | Y | 1. Free<br>2. Distributed<br>3. Highly available<br>4. Master-less replication with robust support for clusters across multiple datacenters |
| 13. | CDB or Constant Database [78] | N | N | Y | N | 1. Free library<br>2. On-disk associative array that maps keys to values, allowing a key to have multiple values<br>3. It can be used as a shared library. |
| 14. | Cloudant [79] | Y | N | N | N | 1. Proprietary<br>2. Distributed database service<br>3. Uses BigCouch [189] and JSON model, at backend |
| 15. | Clusterpoint Database [80] | Y | N | Y | N | 1. Proprietary, but allows download for free<br>2. Distributed JSON/XML database platform<br>3. Transactions are compliant with ACID properties<br>4. Highly available<br>5. Provides sharding and replication<br>6. Uses SQL or JS as query language |
| 16. | Coherence [81] | N | N | Y | N | 1. Proprietary<br>2. In-memory data grid and distributed cache<br>3. Appropriate for systems requiring high scalability and availability keeping latency at lower levels |
| 17. | CouchBase Server [82] | Y | N | Y | N | 1. Free<br>2. Distributed database<br>3. Uses SQL as querying language<br>4. Uses JSON model |
| 18. | CouchDB [83] | Y | N | N | N | 1. Free<br>2. Supports JSON over HTTP/REST<br>3. Provides limited support for ACID transactions<br>4. Supports multi-version concurrency control |
| 19. | CrateIO [84] | Y | N | N | N | 1. Free |
| | | | | | | 2. It is based on Elasticsearch/Lucene ecosystem<br>3. Supports objects that are binary or are also called BLOBs.<br>4. Makes use of SQL syntax for distributed querying of the system in real time |
| 20. | CosmosDB [85] | Y | N | N | N | 1. Proprietary<br>2. Provisioned as Platform-as-a-Service<br>3. Based on DocumentDB |
| 21. | DataStax Enterprise Graph [86] | N | Y | N | N | 1. Proprietary<br>2. Scalable, distributed database<br>3. Allows real-time querying<br>4. Supports Tinkerpop [197]<br>5. It is known to integrate well with Cassandra [77] |
| 22. | DocumentDB [87] | Y | N | N | N | 1. Proprietary<br>2. Provisioned as a database service<br>3. Fully managed version of MongoDB |
| 23. | Dynamo [88] | N | N | Y | N | 1. Proprietary<br>2. Distributed datastore<br>3. Highly available<br>4. Supports incremental scalability, symmetry among nodes, decentralization and it exploits the heterogeneity of the infrastructure it works on. |
| 24. | ElasticSearch [89] | Y | N | N | N | 1. Free<br>2. Supports JSON<br>3. Basically a search engine |
| 25. | etcd [90] | N | N | Y | N | 1. Free<br>2. Supports binary data<br>3. Allows versioning, validation, collections, triggers, clustering, Lucene full text search, ACLS and XQuery Update<br>4. Uses XML over REST/HTTP |
| 26. | eXist [91] | Y | N | N | N | 1. Free<br>2. Supports text, JSON, HTML and XML formats, in addition to binary formats<br>3. XQuery is the provided querying language while XSLT is the corresponding programming language |
| 27. | FoundationDB [92] | N | N | Y | N | 1. Free<br>2. Complies with ACID properties<br>3. Scalable<br>4. Allows replications<br>5. Bindings for Python, C, PHP and Java, in addition to many other programming languages is available |
| 28. | GridGain Systems [93] | N | N | Y | N | 1. Proprietary<br>2. Services and software solutions are provided for systems dealing with big data<br>3. Supports in-memory computing<br>4. Provides improved throughput and reduced latency |
| 29. | GT.M [54] | N | N | Y | N | 1. Free<br>2. Developed for transaction processing<br>3. Supports ACID transactions<br>4. Supports replication and database encrytion |
| 30. | Hazelcast [95] | N | N | Y | N | 1. Free<br>2. MapStore can be defined by the user<br>3. MapStore can be persistent<br>4. High consistency and supports sharing in the form of consistent hashing |
| 31. | HBase [4] | N | N | N | Y | 1. Free<br>2. Distributed database<br>3. It runs on top of Hadoop and provides capabilities similar to that of BigTable.<br>4. It is fault-tolerant for scenarios where a large amount of sparse data is being dealt with. |
| 32. | Hibari [96] | N | N | Y | N | 1. Free<br>2. Distributed big data store<br>3. Highly available<br>4. Strongly consistent |
| 33. | HyperGraphDB [97] | N | Y | N | N | 1. Free<br>2. Schemas are dynamic and flexible<br>3. Knowledge representation and data modeling are efficient<br>4. Non-blocking concurrency<br>5. Appropriate for semantic web and arbitrary graph use cases |
| 34. | HyperTable [98] | N | N | N | Y | 1. Proprietary<br>2. It is based on BigTable<br>3. Massively scalable |
| 35. | IBM Informix [99] | Y | N | N | N | 1. Proprietary<br>2. RDBMS that supports JSON<br>3. Complies with ACID rules<br>4. Supports sharding and replication |
| 36. | IBM Informix C-ISAM [100] | N | N | Y | N | 1. This API complies with Open Standards<br>2. Allows management of data files, which have been organized using B+ indexing<br>3. It is the file storage used by Informix [99]. |
| 37. | Ignite [101] | N | N | Y | N | 1. Free<br>2. Distributed, in-memory computing platform<br>3. Provides caching and processing platform<br>4. Provides support for ACID transactions and MapReduce jobs<br>5. Allows partitioning, clustering and replication<br>6. Highly consistent |
| 38. | InfiniteGraph [102] | N | Y | N | N | 1. Proprietary<br>2. Cloud-enabled<br>3. Distributed<br>4. It is scalable and cross-platform<br>5. It is capable of handling high throughput |
| 39. | InfinityDB [103] | N | N | Y | N | 1. Proprietary<br>2. Completely developed in Java and includes DBMS and database engine<br>3. Based on B-tree architecture<br>3. Provides high performance<br>4. Reduces risks associated with failures |
| 40. | Jackrabbit [104] | Y | N | N | N | 1. Free<br>2. Implementation of Java Content Repository |
| 41. | JanusGraph [105] | N | Y | N | N | 1. Free<br>2. Distributed<br>3. Scalable and integrates well with backend databases like HBase [4], Cassandra [77], BigTable [75] and BerkleyDB [73]<br>4. Integrates well with platforms like Giraph [144], Spark [8] and Hadoop [12]<br>5. Provides support for full text search by external integration with Solr [137] and Elasticsearch [89] |
| 42. | KAI [106] | N | N | Y | N | 1. Free<br>2. Scalable |
| | | | | | | 3. Highly fault-tolerant<br>4. Provides low latency<br>5. Used for social networks and web repositories |
| 43. | LevelDB [108] | N | N | Y | N | 1. Free<br>2. Maintains byte arrays for storing key and value pairs.<br>3. Data compression is supported by means of Snappy<br>4. Supports forward/backward iteration and batch writing<br>5. Used as a library |
| 44. | Lightening Memory-Mapped Database (LMDB) [109] | N | N | Y | N | 1. Free<br>2. Embedded database<br>3. High performance<br>4. Provides API bindings for many programming languages.<br>5. Employs multi-version concurrency control is offers high levels of reliability |
| 45. | Lotus Domino [110] | Y | N | N | N | 1. Proprietary<br>2. It is a multi-value database [190] |
| 46. | Marklogic [111] | Y | Y | N | N | 1. Free<br>2. Supports XML, JSON and RDF triples<br>3. Distributed<br>4. Provides high availability, full-text search, ACID compliance and security |
| 47. | Memcached [112] | N | N | Y | N | 1. Free<br>2. Memory caching system that is general purpose and distributed<br>3. Scalable architecture<br>4. Supports sharding |
| 48. | MemcacheDB [113] | N | N | Y | N | 1. Free<br>2. A version of memcached that has persistence<br>3. It is a memory caching system that is distributed and general purpose.<br>4. Development has halted on this solution. |
| 49. | Microsoft SQL Server [114] | N | Y | N | N | 1. Proprietary<br>2. Typically used for modeling many-to-many relationships between data<br>3. Integration of relationships are done into Transact-SQL and the foundation DBMS is SQL Server |
| 50. | MongoDB [6] | Y | N | N | N | 1. Free<br>2. Supports BSON or binary JSON<br>3. Allows replication and sharding |
| 51. | MUMP Database [115] | Y | N | N | N | 1. Proprietary<br>2. MUMPS is a programming language with inbuilt database<br>3. Used for applications related to health sector |
| 52. | Neo4j [116] | N | Y | N | N | 1. Free<br>2. Can be used for ACID transactions<br>3. Supports clustering and high availability<br>4. Provides complete administrative support<br>5. Provides inbuilt REST API for interface with other programming languages |
| 53. | NoSQLz [117] | N | N | Y | N | 1. Proprietary<br>2. Complies with ACID properties<br>3. Allows CRUD (Create, Read, Update, Delete) operations<br>4. Easy to implement |
| 54. | ObjectDatabase++ [118] | Y | N | N | N | 1. Proprietary |
| | | | | | | 2. Binary<br>3. Structure of native C++ class |
| 55. | OpenLink Virtuoso [119] | Y | Y | Y | N | 1. Proprietary<br>2. Hybrid of database engine and middleware<br>3. High performance and secure<br>4. Supports SQL [23] and SPARQL [196] for performing operations on SQL tables and RDF<br>5. JSON, XML and CSV document types are supported |
| 56. | Oracle NoSQL Database [120] | N | N | Y | N | 1. Proprietary<br>2. Supports horizontal scalability and transparent load balancing<br>3. Supports replication and sharding<br>4. It is highly available and fault-tolerant |
| 57. | Oracle Spatial and Graph [121] | N | Y | N | N | 1. Proprietary<br>2. Capable for handling RDF and property graphs |
| 58. | OrientDB [122] | Y | Y | N | N | 1. Free<br>2. Supports JSON over HTTP<br>3. Supports SQL-type language use<br>4. Can be used for ACID transactions<br>5. Supports sharding, multi-master replication, security features and schema-less modes |
| 59. | PostgreSQL [123] | Y | N | N | N | 1. Free<br>2. Supports JSONB, JSON function and JSON store<br>3. Supports HStore 2 and HStore [192] |
| 60. | Project Voldemort [124] | N | N | Y | N | 1. Free<br>2. Supports horizontal scalability<br>3. Availability is high for read/write operations<br>4. Fault recovery is transparent<br>5. Supports automatic partitioning and replication<br>6. Considered appropriate for applications with read-intensive operations |
| 61. | Qizx [125] | Y | N | N | N | 1. Proprietary<br>2. Distributed XML database<br>3. Supports text, JSON and binaries<br>4. Provides integrated full text search |
| 62. | RavenDB [126] | Y | N | N | N | 1. Free<br>2. Fully transactional and high performance<br>3. Highly available<br>4. Multi-platform and easy to use<br>5. Multi-model architecture that allows it to work well with SQL systems |
| 63. | Redis [127] | N | N | Y | N | 1. Free<br>2. Read/write access is efficient<br>3. Fault-tolerant<br>4. Supports automatic partitioning<br>5. Appropriate for applications involving structured strings |
| 64. | RethinkDB [128] | Y | N | N | N | 1. Free<br>2. Distributed database<br>3. Supports JSON<br>4. Provides sharding and replication |
| 65. | Riak [129] | N | N | Y | N | 1. Free<br>2. Highly available and fault-tolerant<br>3. Highly scalable and simple to operate<br>4. Cloud storage and enterprise versions of Riak are also available<br>5. It supports automatic data distribution and |
| | | | | | | replication for resilience and improved performance. |
| 66. | RocketU2 [130] | Y | N | N | N | 1. Proprietary<br>2. Provides dynamic support<br>3. Scalable<br>4. Reliable and efficient<br>5. Appropriate for business information management |
| 67. | RocksDB [131] | N | N | Y | N | 1. Free<br>2. Embedded database that assures high performance<br>3. Supports all the features of LevelDB [108]. In addition, it also supports geospatial indexing, universal compaction, column families and transactions. |
| 68. | SAP HANA [132] | Y | Y | N | N | 1. Proprietary<br>2. Supports JSON only<br>3. Can be used for ACID transactions |
| 69. | Scalaris [133] | N | N | Y | N | 1. Free<br>2. Highly available and fault-tolerant<br>3. Massively scalable<br>4. Consistent<br>5. Self-managing<br>6. Minimal maintenance overhead<br>7. Considered appropriate for applications that are read/write intensive |
| 70. | ScyllaDB [134] | N | N | N | Y | 1. Free<br>2. Distributed<br>3. Designed for integration with Cassandra for reducing latency and improving throughput<br>4. It supports Thrift and CQL, protocols also supported by Cassandra |
| 71. | Sedna [135] | Y | N | N | N | 1. Free<br>2. XML database |
| 72. | SimpleDB [136] | Y | N | N | N | 1. Proprietary<br>2. Distributed database<br>3. Used as web service in concert with Amazon EC2 [193] and S3 [194]<br>4. Provides availability and partition tolerance |
| 73. | Solr [137] | Y | N | N | N | 1. Free<br>2. Search engine written in Java<br>3. Supports real-time indexing, full text search, database integration, dynamic clustering and rich document handling<br>4. Provides index replication and distributed search<br>5. Scalable and fault tolerant |
| 74. | Sparksee [138] | N | Y | N | N | 1. Proprietary<br>2. Scalable<br>3. High performance<br>4. First graph database for mobiles<br>5. Bindings available for C++, C#, Objective C, Python and Java |
| 75. | Sqrrl [139] | N | Y | N | Y | 1. Proprietary<br>2. Distributed<br>3. Mass-scalable<br>4. Real-time database<br>5. Provides cell-level security |
| 76. | Tarantool [140] | N | N | Y | N | 1. Free<br>2. Provides crash resistance with the help of |
| | | | | | | maintenance of write ahead logs<br>3. Can be integrated with other applications and frameworks written in different programming languages. |
| 77. | TokuMX [141] | Y | N | N | N | 1. Free<br>2. Version of MongoDB [6]<br>3. Supports fractal tree indexing [195] |
| 78. | TerraStore [143] | Y | N | N | N | 1. Free<br>2. In-memory storage<br>3. Dynamic cluster configuration<br>4. Persistent<br>5. Supports load balancing and automatic data redistribution<br>6. Used for structured big data |
| 79. | Tokyo Cabinet and Kyoto Cabinet [142] | N | N | Y | N | 1. Free<br>2. Provides two libraries for database management<br>3. Storage is done via hash tables and B+ trees<br>4. Provides limited support for transactions |
| 80. | XAP [146] | N | N | Y | N | 1. Proprietary<br>2. Software platform for in-memory computing<br>3. Appropriate use cases for this solution include real-time analytics and transaction processing requiring low latency and high performance levels. |

## 4.0 Big Data File Formats: Storing Data in Hadoop Ecosystem

Hadoop [12] offers one of the most cost-effective and efficient ways to store data in huge amounts. Moreover, structured, semi-structured and unstructured data types can be stored and then, processed using tools like Pig [147], Hive [148] and Spark [8] to gain the results required for any future analysis or visualization. Hadoop allows the user to store data on its repository in several ways. Some of the commonly available and understandable working formats include XML [149], CSV [150] and JSON [151].

Although, JSON, XML and CSV are human-readable formats, but they are not the best way, to store data, on a Hadoop cluster. In fact, in some cases, storage of data in such raw formats may prove to be highly inefficient. Moreover, parallel storage is not possible for data stored in such formats. In view of the fact that storage efficiency and parallelism are the two leading advantages of using Hadoop, the use of raw file formats may just defeat the whole purpose.

CERN [152] had chosen four candidate technologies, ORC [153], Parquet [154], Kudu [155] and Avro [156] for this purpose. However, we have included other data file formats like Arrow [157] and text-based formats to this discussion for comprehensibility. Fig. 3 shows the classification and hierarchy of big data file formats. Each of these file formats has a set of advantages and disadvantages to its name. This section explores these facets of big data file formats and shall provide a discussion on the criteria that must be used for selecting a file format.

### 4.1 Text-Based Formats

Simplest example of such a data file format is entries stored line-by-line terminated by a newline character or carriage return. In order to compress data, a file-level compression codec like BZIP2 [] needs to be used. Three formats namely text or CSV [150], JSON [151] and XML [149] are available under this category.

4.1.1 Plain Text

Data stored in this format is mostly formatted in such a manner that fields either has fixed width or a delimiter separated entries, as is the case of CSV in which entries are separated using commas.

4.1.2 XML

It is possible to use external schema definitions in XML. However, the performance of serialization and deserialization is usually poor.

4.1.3 JSON

JavaScript Object Notation (JSON) [158] is more performance effective than XML, but the problem with serialization and deserialization performance exists.

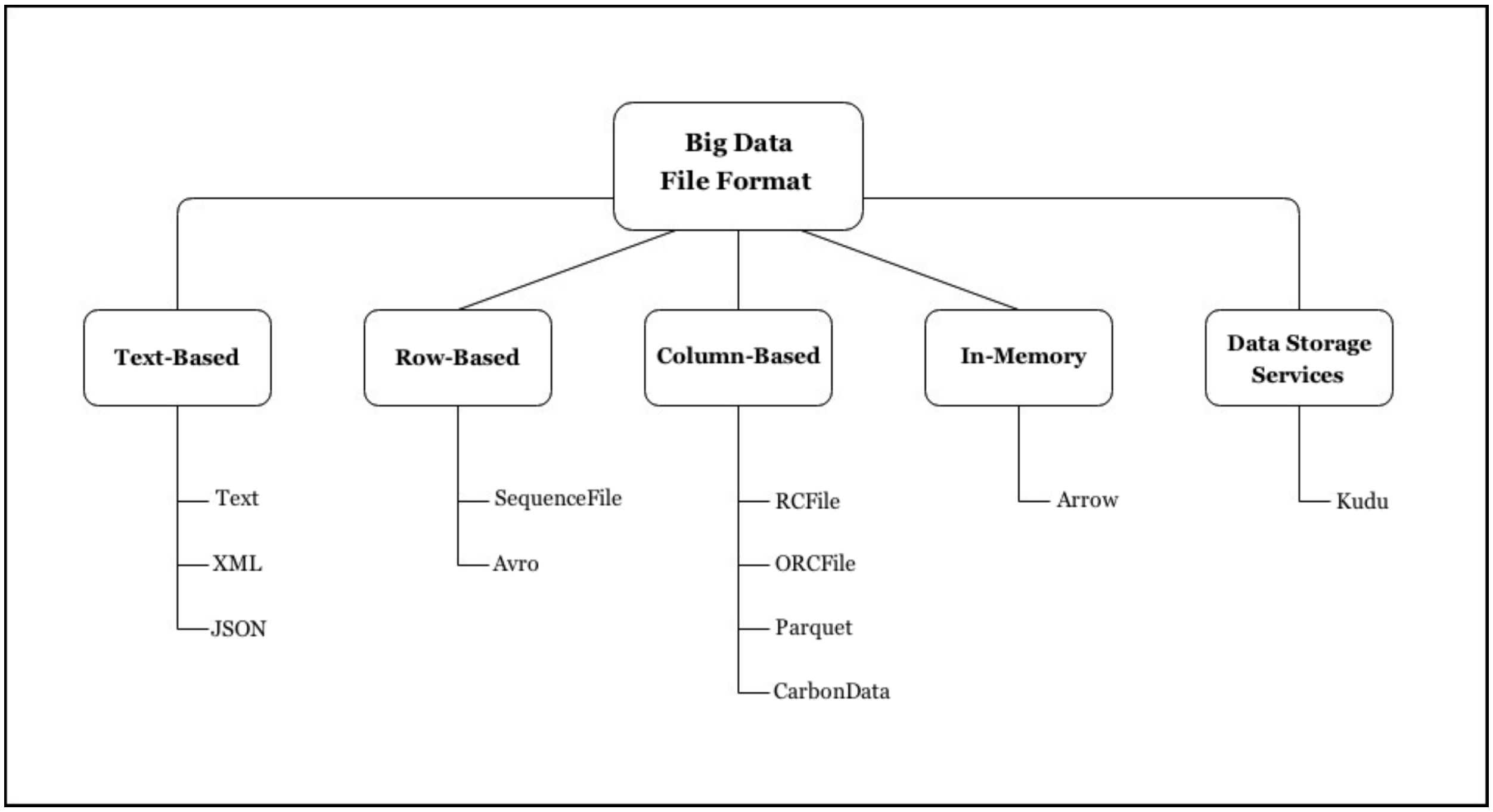

Fig. 3. Classification of Big Data File Formats

**4.2 Row-Based Formats**

4.2.1 SequenceFile

This file format is supported as part of the Hadoop framework as part of which a large file has container of binary key-value pairs for storing multiple files of small size. Therefore, the key-value pairs corresponding to records are encoded. The integration of SequenceFile [159] data file format is smooth with Hadoop in view of the fact that the former was developed for the latter.

4.2.2 Avro

Apace Avro [156] is used for compact binary format as a data serialization standard. Typically, it is used to store persistent data related to communication protocols and HDFS [5]. One of the major advantages of using Apache Avro is high ingestion performance, which is attributed to fast and lightweight deserialization and serialization, provided by the same. It is important to mention that Avro does not have an internal index. However, the directory-based partitioning technique available in HDFS can be used for facilitating random access of data. Data compression algorithms supported by Apache Avro include DEFLATE [160] and Snappy [161].

There are other serialization and deserialization frameworks like Thrift and Protocol buffers [162] that stand in competition with Avro. However, Avro is a built-in component of Hadoop while these frameworks are external. Besides this, schema definition in Avro is done using JSON whereas Thrift and Protocol buffers depend on Interface Definition Languages (IDLs) [162] for defining the schema.

**4.3 Column-Based Formats**

4.3.1 Record Columnar (RC) File Format

This is the most primitive columnar record format that was created as part of the Apache Hive project. RCFile [163] is a binary format similar to SequenceFile that assures high compression for operations that involve multiple rows. Columns are stored as a record in a columnar fashion by creating row splits. Vertical partitions are created on row splits in a columnar manner. The metadata is stored for each row split as the key and the corresponding data is kept as its value.

4.3.2 Optimized Row Columnar (ORC) File Format

Optimized Row Columnar [153] is similar to Parquet and RCFile in the sense that all these three data file formats exist within the Hadoop framework. The read performance is better. However, writes are slower than average. Besides this, the encoding and compression capabilities of this file format are better than that of its counterparts, with ORC supporting Zlib [164] and Snappy [161].

4.3.3 Parquet

Apache Parquet [154] is a data serialization standard, which is column-oriented and known to improve the efficiency significantly. This data format also includes optimizations like compression on series of values that belong to the same column resulting in improved compaction ratios. Besides this, encodings like bit packing, dictionary and run length encoding are additional optimizations available. In order to compress data, Snappy [161] and GZip [165] algorithms are supported.

4.3.4 CarbonData

This data format was developed by Huawei to manage existing shortcomings in already available formats. CarbonData [166] is a relatively new data file format that allows developers to reap the benefits of column-oriented storage while also providing support for handling random access queries. Data is grouped into blocklets, which is stored alongside other information about the data like schema, indices and offsets, in addition to others. Metadata is stored in headers and footers, which offers significant performance optimization during scanning and processing of subsequent queries.

**4.4 In-Memory Formats**

Apache Arrow [157] is a platform for development of applications using in-memory data. Moreover, it works across languages, which makes it a standard for columnar memory format, enabling support for hierarchical as well as flat data. The data is organized to provide high performance analytics on modern hardware. Interprocess communication and streaming works on zero-copy or no deserialization and serialization. Besides this, it provides many computational libraries for advanced complexities.

**4.5 Data Storage Services**

Kudu [155] is a storage system that ensures scalability and is based on the concept of distributed storage. Data is stored inside tables. Moreover, this data format achieves optimized trade-off between performance and speed of ingestion, which is attributed to the columnar data organization and maintenance of index. Data compression algorithms supported by Apache Kudu include LZ4 [167], Zlib [164] and Snappy [161].

### *4.6 Comparison of Big Data File Formats*

The actual storage of data on a file system is determined by the chosen data file format, which is a crucial choice to make in view of the fact that it plays a significant role in optimizing system storage. The decision of choosing a file format for an application depends on the use-case or the algorithm being used for data processing. With that said, the chosen file format must fulfill some basic requirements to be deemed appropriate for big data systems.

Firstly, the chosen big data file format must be expressive and well defined. Secondly, it should have diverse handling capabilities as far as supported data structures are concerned. Some of the basic structures that must be supported include structs, maps, numbers, strings, records and arrays, to name a few. Finally, the big data file format must be binary, simple and provide support for compression.

One of the biggest bottleneck issues in HDFS-related applications that make use of technologies like Spark and Hadoop include reduction in data read and write times. Issues like storage constraints, evolving schemas and big data further complicate the system requirements. In order to mitigate these issues across application domains and problem scenarios, several big data file formats have come into being. The use of an appropriate file format can benefit the system in the following ways –

1. Read time is reduced.
2. Write time is reduced.
3. The files can be split, which in other words means that there is no longer the need to read the whole file for

retrieving a smaller sub-section of it.

4. There is support for schema evolution and schema can be changed on request depending upon the changing needs of the system.
5. There is availability of advanced compression codecs to ensure that files can be compressed without losing the advantages offered by the base format.

In view of the above-mentioned advantages, choosing the right data file format can optimize system performance substantially. However, a plethora of options are available in this regard. While some file formats are developed for general use, there are some others that offer optimization benefits to specific applications or improve specific characteristics. This makes a comparison of the file formats essential for facilitating the decision of which data file format is best suited for an application. Table 3 summarizes the advantages, disadvantages and typical use cases for the different data file formats discussed in the previous sections.

Table 3: Types of Big Data File Formats

| Class | Data File Format | Advantages | Disadvantages | Matching Uses |
|---|---|---|---|---|
| Text-Based Data File Formats | Text<br>XML [149]<br>JSON [151] | 1. Light weight | 1. Read is slow.<br>2. Write is slow.<br>3. Space is wasted because of column headers that are not required.<br>4. Compressed files cannot be split, which leads to huge maps. Moreover, block-level compression is not supported. | Appropriate for starting use of structured data on HDFS. Also, CSV files are used in cases where data needs to be extracted from Hadoop and bulk-loaded into database. |
| Row-Based Data File Format | SequenceFile [159] | 1. This file format is compact in comparison with Text files.<br>2. The file format supports optional compression.<br>3. It supports parallel processing.<br>4. An enormous number of small-sized files can be stored in a ‘container,’ provided for the same purpose.<br>5. One of the biggest advantages of this data file format is the support for block-level compression that allows file compression while allowing file splitting for multiple tasks. | 1. This file format is not preferred for tools like Hive.<br>2. The append functionality of the file format just as good and comparable to other file formats.<br>3. The file format lacks support for multiple languages. | If intermediate data, which is generated between jobs, needs to be saved, then SequenceFile data file format is used. |
| | Avro [156] | 1. The size of serialized data is smallest.<br>2. It offers block-level compression, allowing file splitting at the same time.<br>3. This file format maintains the structure of the object.<br>4. Even if the schema has changed, Avro allows reading of old data.<br>5. Schema definitions are written in JSON. Therefore, development is considerably simplified in programming languages that possess JSON libraries. | 1. Reading and writing processes need schema definition. | Avro is considered best for cases where schema evolution is a key requirement. If the schema is expected to change over time, then Avro is preferred. In fact, Avro is preferred for all Hadoop-based use cases. |
| Column-Based | RCFile [163] | RCFile offer typical benefits | 1. There is no support for schema | If the use case involves |
| Data File Format | | associated with columnar databases, which include –<br>1. It offers good compression.<br>2. The query performance is better than that for row-oriented databases. | evolution.<br>2. The write process is slower. | tables that possess many columns and the application requires frequent use of specific columns, then RCFile is the preferred data format. |
| | ORCFile [153] | 1. The compression capabilities of ORCFile are better than that of RCFile.<br>2. Query processing is also improved when compared to RCFile. | 1. There is no support for schema evolution.<br>2. ORCFile format is not well supported by Apache Impala. | ORCFile and Parquet are used in scenarios where query performance is crucial. However, it has been found that Parquet when used with SparkQL show best improvements in query performance. |
| | Parquet [154] | 1. As is the case with ORCFile format, compression and query performance are good. Moreover, in cases where specific columns are being queried, this data file format is particularly effective.<br>2. The read performance is good.<br>3. Schema evolution, in this case, is better than ORCFile format as columns can be added at the end.<br>4. Storage optimizations are remarkable and it offers file-level as well as block-level compression. | 1. Writes are computationally intensive.<br>2. If the application requires rows of data, then like all columnar databases, Parquet may not be performance-effective in view of the network activity overheads involved. | |
| | CarbonData [166] | 1. It supports update and delete operations, which is crucial for many workflows.<br>2. CarbonData offers optimizations like bucketing and multi-layer indexing. Therefore, joining two files is more performance-effective and queries are speedier than ever. | 1. CarbonData does not support ACID.<br>2. The size of compressed files is larger than those obtained with ORC and Parquet.<br>3. CarbonData is a relatively new data file format with many technologies like Athena [38] and Presto [94] not supporting it yet. | CarbonData can be used for variable analytical workloads with typical use cases involving interactive or detailed queries and queries involving real-time ingestion and aggregating/OLAP BI. |
| In-Memory Data File Format | Arrow [157] | 1. This format enables systems to handle big datasets.<br>2. Same memory can be shared by multiple applications by means of a common data access layer.<br>3. This file format is optimized for parallel processing and data locality. Moreover, it has been developed for Single Instruction Multiple Data (SIMD).<br>4. It allows processing of scan as well as random access workloads.<br>5. The overheads associated with streaming messages and RPC are significantly reduced.<br>6. Apache Arrow can be used | 1. As of now, predicate push down implementation is left to the engine. Although, Apache Arrow is expected to reusable, fast vectorized operations, but efforts in this direction are yet to take shape. | Apache Arrow is best suited for vectorized processing that involve Single Instruction on Multiple Data (SIMD). |
| | | with GPUs.<br>7. The columnar format provided by Apache Arrow is 'fully shredded' and supports nested and flat schemas. | | |
| Data Storage Services | Kudu [155] | 1. OLAP workloads can be processed quickly with Kudu.<br>2. Kudu can be seamlessly integrated with Spark, Hadoop and its ecosystem.<br>3. It can be tightly integrated with Apache Impala, which is an effective alternative to Parquet with HDFS.<br>4. The system is highly available.<br>5. The data model provided is structured.<br>6. The management and administration of Kudu is simple.<br>7. The consistency model is flexible and strong.<br>8. Random and sequential workloads can be simultaneously executed with high performance. | 1. There are no data restore or backup options inbuilt in Kudu.<br>2. There are some security limitations like authorization available only at the system level and no inbuilt support for data encryption.<br>3. Kudu does not support automatic partitioning and repartitioning of data.<br>4. There are other schema-level limitations like lack of support for secondary indexes and multi-row transactions. | Kudu has been created for applications centered on time series data and for applications like online reporting and machine data analytics. |

It can be inferred from the comparison table that even though text-based formats are simple and lightweight, they present a host of drawbacks that can affect the performance of the system considerably. In order to overcome the limitations of text-based formats, Hadoop has inbuilt data file formats. The first of these formats is a row-based data file format called SequenceFile [159]. Other data file formats that are based on SequenceFile and are included in the Hadoop ecosystem include MapFile, SetFile and BloomMapFile [168]. These file formats are designed for specialized use cases. SequenceFile only supports Java, which makes it language-dependent, and does not support versioning. As a result, Avro [156], which overpowers SequenceFile, has come up as the most popular row-based data file format. Understandably, Avro is the best option among row-based data file formats.

In case of row-oriented file formats, contiguous storage of rows is done in the file. On the other hand, in case of column-oriented formats, rows are split and values for a row-split are stored column–wise. For example, the values for the first column of a row slit are stored first and so on. Therefore, columns not required for a query's processing can be omitted during data access. In other words, row-oriented formats are best suited for cases in which fewer rows are to be read, but many columns for the row are required. On the other hand, if a small number of columns are required, then column-oriented file formats are a better fit.

It is important to note that column-oriented data file formats require a buffer for row splitting as it works with more than one row at a time. Moreover, it is not possible to control the writing process. If the process fails, it is not possible to recover the current file. This is the reason why column-oriented formats are not used for streaming writes. On the other hand, the use of Avro allows read up to the point until which sync had occurred. Flume makes use of row-oriented formats because of this property [191].

Systems are developed in such a manner that seeks to a disk are kept to a bare minimum and thus, latency is reduced. This is an efficient storage mechanism for transactional workloads for which data is written row-wise. For analytical workloads, large number of rows needs to be accessed at the same time. However, a subset of columns may have to be read for processing a query. In such scenarios, row-oriented format is inefficient considering the fact that all the columns of multiple rows will have to be read even if all these columns are not required. The use of column-oriented format is expected to reduce the number of seeks, improving query performance. Although, writes are slower in this case, but for analytical workloads, this is expected to work well as the number of reads are expected to outnumber writes.

RCFile [163] is the most primitive column-based data file format and ORCFile is an optimized version of the same. Although, ORCFile [153] is considered apt for applications with ACID transactions and offers fast access with its indexing feature, Parquet is promoted by Cloudera [169] and is known to perform best with Spark [8]. The most advanced and recent file format in this category is CarbonData [166], but owing to its newer status, its compatibility with technologies is questionable. This makes Parquet the most popular column-based data file format. Arrow and Kudu [155] support columnar representation with the difference that Arrow supports 'in-memory' storage while Kudu provisions storage that is mutable on disk as against Parquet format, which is immutable on disk and on disk storage.

The tradeoffs for using immutable data file format (Parquet [154]) include higher throughput for write, easier concurrent sharing, access and replication, and no overheads related to operations. However, for making any modifications, dataset needs to be rewritten. Mutable (Kudu [155]) allows higher flexibility in the compromise between speeds of making updates and reads. The latency for short access is lower owing to quorum replication and primary key indexing. Moreover, the semantics are similar to that of databases. With that said, a separate daemon is required for management.

When comparing Parquet's on-disk storage with transient in-memory storage of Arrow [157], it is important to note that the former allows multiple query access with priority given to I/O reduction, but CPU throughput must be good. However, on-disk storage is deemed appropriate for streaming access only. On the other hand, in-memory storage supports streaming, as well as random access, and is focused towards execution of a single query. In this case, CPU throughput is prioritized even though I/O throughput must also be good. Some projects make use of Arrow and Parquet together. Pandas [170] is one example of such a usage. The data frame from Pandas can be saved onto Parquet, which can then be read onto Arrow. The ability of Pandas to work on columns of Arrow, allows it to easily integrate with Spark.

## 5.0 Future Trends in Big Data Storage Technology

Technologies used for big data storage have matured over time to give way to next generation technologies like blockchain, which showed a remarkable rise in popularity and usage in a broad collection of domains [171]. As is the case with any technology that is used across domains, several challenges and issues have been identified in its applications to real world scenarios. One of the most profound challenges facing this technology is scalability [172]. However, the capability of this technology to take Internet to its next phase by creating a decentralized web [173] makes it worth a discussion.

One of the fundamental components of the decentralized web is decentralized storage [174]. In addition to blockchain, some other technologies like Internet of Things (IoT) [175] and artificial intelligence [176] are also testing grounds of existing storage solutions and their capabilities. The rise of Internet of Things (IoT) and its expected coverage of 20 billion connected devices [177] by the year 2020 are expected to require acquisition, storage, management and processing of huge data reserves.

The demand from storage solutions is increasing in view of the challenges posed by data management for connected devices, data sharing and need for personalization in applications [178]. There is a shift of approach towards data-intensive applications particularly in view of the dependency of companies on heaps of data. However, this data is stored in centralized data centers, which leads to multiple security breaches and issues [179]. This argument supports the use of decentralized data storage solutions.

However, building decentralized applications on top of blockchain technologies offers its set of independent challenges. For instance, data management and storage are not supported by solutions like Ethereum [180]. Therefore, when these solutions are forced to perform beyond their capacity, they consume larger space and more time. The vision behind decentralized storage is to amalgamate the features of blockchain technology with solutions that can cater to the practical demands of big data storage. It is evident from the name itself that distributed storage divides data into smaller data units and distributes it to different nodes where it is stored. This characteristic can be compared to the distributed ledger technology [181] that is commonly associated with blockchain.

Presently, even cloud-databases are too centralized and easy for hackers to target [182]. Moreover, the points of failure are highly obvious, which makes them all the more, easier to attack. Power outage is one such point of failure [183]. On the contrary, a decentralized system is devoid of such issues in view of the fact that the nodes are geographically distant and physically unrelated. Therefore, an outage or attack on a point of failure is expected to affect only the node concerned. The system will overall remain unaffected, as the other nodes of the system will continue to function as usual. Apart from making the system reliable and available, decentralization also contributes to the scalability and performance of the system [184].

While drawing parallels between decentralized storage and blockchain, it is important to realize that storage on blockchain remains a significant issue [174]. As more and more transactions take place on the blockchain, scalability might become complicated. Therefore, storing big data is certainly a questionable domain from the blockchain perspective. In order to mitigate the challenges mentioned above, techniques like swarming [185] and sharding [186] are being used. Partitioning of databases along logical lines is referred to as sharding. The decentralized model ensures storage of shards together. Moreover, a unique partition key is used by a dedicated decentralized application to access the shards. Besides this, swarming is used to enable collective storage of shards. Data is stored and managed by creating a large group of nodes, which is called a swarm. This group of nodes is similar to the network of nodes created for blockchain.

The most profound advantages of swarming are reduced latency and increased speed [185] considering the fact that data is typically retrieved from the fastest and nearest nodes. In addition to this, the nodes are geographically distant, which makes the system scalable and reliable. Another important aspect of decentralized storage from the customer's point of view is that the customer will have the choice to buy node-level storage from different vendors as against a compulsive single vendor. This can be seen as a major step towards facilitating adoption of this technology in real-world scenarios. It can be concluded that decentralized storage is sure to offer a scalable, efficient and secure solution for the future of big data storage.

## 6.0 Conclusion

Design and development of a big data application that can resolve real world problems and prove to be a viable solution is dependent on the base technologies chosen for the creation of a heterogeneous storage and computing environment. The scope of this research paper is limited to comparison and analysis of storage solutions available for big data systems. There are two facets of storage infrastructures in the big data context. The first facet deals with data storage and management by a dedicated storage infrastructure. In view of specific storage challenges posed by big data like scalability, availability, integration and security, traditional systems are deemed incapable to handle the existing data scenario.

NoSQL has proven to be a viable solution to the big data problem. Few of the significant features of NoSQL that prove the feasibility of its usage for big data storage and management include aspects like easily scalable systems, flexible data modeling, high availability and provisioning of required performance considering the fact that most modern systems handle static as well as real-time data. Features of NoSQL like dynamic schema, auto-sharding, automatic replication and integrated caching abilities mitigate the challenges posed by big data to traditional systems.

Understandably, data is the heart of the system and modeling data is the most crucial design activity for optimum system performance and functionality. Different research works have classified available NoSQL solutions on the basis of supported data models. However, the most accepted classification categorizes NoSQL solutions on the basis of four data models namely document-oriented, graph, key-value and wide-column. The research paper analyzes the benefits and shortcomings of each of these data models to provide an analysis of the use cases that are most appropriate for each data model and the ones that are not. The market is flooded with solutions and technologies that provision a combination of customizable storage, acquisition, processing and visualization [188] facilities to the developer.

This decision is based of many factors that can be technical and non-technical in nature. Once the right data model is determined for a big data problem, a solution that supports the data model along with the desired features as per the requirements of the big data system need to be found. This research paper provides feature analysis of 80 NoSQL solutions to facilitate decision making in this regard. With that said, benchmarking of these solutions and their performance analysis for a complete quantitative comparison can be performed in the future. Moreover, it has been found that many solutions support multiple data models, making the classification categories insufficient.

It is suggested that hybrid data model categories must be used for discrete classification of solutions to make the decision making process simpler for use-cases that include multi-dimensional data, which might require multiple base data models to design. The second facet of big data storage hails from the fact that different technologies in a heterogeneous technological environment need to share data to operate. The most optimized method for such data sharing is the use of common data file formats. This research paper classifies available big data file formats into five categories namely text-based, row-based, column-based, in-memory and data storage services. Text-based formats have lost utility in the era of big data. However, JSON and XML format usage is common considering the fact that they are lightweight and human readable.

Hadoop provides better ways to store and format data on files. Closest to the relational way of storing data is row-based format. SequenceFile and Avro belong to this category, of which former is the most primitive row-based format provided to users. Other specialized row-based formats that use SequenceFile at their base are also available in Hadoop. These include MapFile, SetFile and BloomMapFile. Avro is the most commonly used row-based format and is preferred for applications that may require all or a majority of the columns.

In most big data scenarios, column-based file formats are known to perform better than their row-based counterparts, as most queries require retrieval of a few columns for multiple rows. Three types of column-based formats are available namely mutable, on-disk storage (Kudu), immutable, on-disk storage (Parquet) and in-memory storage (Arrow). It is important to mention that even though RCFile and ORCFile are also available, Parquet is the most popular column-based format for that category. Some systems use a combination of these formats depending on the requirements of the application. Case study-based performance comparison for quantitative analysis can be performed as future work in this domain.

Besides this, technological decisions are also driven by technical expertise. It has been found that developers accustomed of working on a technology are expected to choose it over solutions that might provide better performance. However, owing to project requirements, they might have to switch to mightier solution, but this wastes development time and effort. Efforts must be made to alleviate such issues. The world of big data storage is drifting from centralized storage to its decentralized counterpart and decentralized storage with blockchain has been identified as the future of big data technology. However, such a synergistic use is faced with several challenges like management of space and time constraints, which must be rigorously dealt with in future work to determine practical feasibility and applicability of such a usage.

## Acknowledgements

This work was supported by a grant from "Young Faculty Research Fellowship" under Visvesvaraya PhD Scheme for Electronics and IT, Department of Electronics & Information Technology (DeitY), Ministry of Communications & IT, Government of India.